\documentstyle[aps, epsf,prl]{revtex}
\begin{document} 
\begin{flushright} {OITS 631}\\
July 1997
\end{flushright}
\vspace*{1cm}

\begin{center}  {\large {\bf Universal Behavior of
Multiplicity Differences\\ in Quark-Hadron Phase Transition}}
\vskip .75cm
 {\bf  RUDOLPH C. HWA }
\vskip.5cm
 {Institute of Theoretical Science and Department of Physics\\ University of
Oregon, Eugene, OR 97403-5203, USA\\E-mail: hwa@oregon.uoregon.edu}
\end{center}
\begin{abstract}
The scaling behavior of factorial moments of the differences in
multiplicities between well separated bins in heavy-ion collisions is
proposed as a probe of quark-hadron phase transition.  The method
takes into account some of the physical features of nuclear collisions
that cause some difficulty in the application of the usual method.  It
is shown in the Ginzburg-Landau theory that a numerical value
$\gamma$ of the scaling exponent can be determined independent of
the parameters in the problem.  The universality of $\gamma$
characterizes quark-hadron phase transition, and can be tested
directly by appropriately analyzed data.
\end{abstract}

\section{Introduction}

The study of multiplicity fluctuations as a phenomenological
manifestation of quark-hadron phase transition (PT) has been
pursued in recent years in the framework of the Ginzburg-Landau (GL)
formalism for both the second-order \cite{1}-\cite{3} and first-order
\cite{4}-\cite{6} PT.  In both cases scaling behaviors of the factorial
moments have been found and are characterized by scaling exponents
$\nu$.  The value of $\nu$ for second order PT is independent of the
details of the GL parameters.  It therefore provides a distinctive
signature for the existence of quark-gluon plasma, if its transition to
hadrons is of the second order \cite{7}.

The experimental verification of $\nu$ has not been carried out so
far in heavy-ion collisions, although it has been checked to a high
degree of accuracy in quantum optics  \cite{8}.  There are a number of
reasons for the difficulties, which will be described below.  The aim of
this paper is to devise a method to circumvent the obstacles that
stand in the way of extracting the signal from the experimental data. 
In so doing we also broaden the scope of the analysis to include
aspects of wavelets and correlations, in addition to incorporating some
evolutionary properties of heavy-ion collisions that are particularly
relevant to hadron production during PT.

Let us now examine the difficulties in analyzing multiplicity
fluctuations in heavy-ion collisions.  The factorial moments that have
been suggested to quantify the fluctuations are defined by \cite{9}
\begin{eqnarray}
F_q = {\left< n(n-1) \cdots (n-q+1) \right> 
\over \left<n\right>^q}
\quad ,
\label{1}
\end{eqnarray}
where the averages are performed over a distribution, $P_n$, of the
multiplicities $n$ in a bin of size $\delta$.  Note that $n$ must be
${\geq q}$ in order for an event to contribute to $F_q$, for which $q$
is usually an integer ranging from 2 to 5.  As $\delta$ is decreased, the
average multiplicity $\left<n\right>$ in a bin decreases, and may
become $\ll q$ in a hadronic collision.   Thus $F_q$ measures the tail
end of the distribution $P_n$, where $n \gg \left<n\right>$, i.e. large
multiplicity fluctuations.  Intermittency refers to the power-law
dependence of $F_q$ on $\delta$ \cite{9}
\begin{eqnarray}
F_q \propto \delta ^{-\varphi _q} \quad ,
\label{2}
\end{eqnarray} 
a behavior that has been found to be ubiquitous in hadronic and
leptonic processes \cite{10}.  However, in nuclear collisions the
situation is very different.  

In the first place the average multiplicity in an event in nuclear collisions
is so high that even in bins of small $\delta$ the values of $n$ are large
compared to the order $q$ in (\ref{1}) that has been examined
experimentally.  Thus unlike hadronic and leptonic processes the existent
nuclear data have not been analyzed to study large multiplicity
fluctuations, for which $q$ must be increased to values $\gg
\left<n\right>$ in small bins.  The moments $F_q$ for $q = 3, 4$, and $5$ that have been
determined are dominated by contributions from the lower-order
correlations.  This point has been emphasized by Sarcevic and
collaborators \cite{11,12}, where the nuclear data on the cumulants
$K_q$ are shown to be consistent with zero for $q \geq 3$.  Whether
$K_q$ acquire nonvanishing values at smaller $\delta$ (so that
$\left<n\right>\ll q$) is not known.  Thus until the experiments can be
improved to render the analysis at very small $\delta$ feasible, a new
method must be devised to circumvent this difficulty of extracting
dynamical information at medium values of $q$.  Our strategy is to
consider the distribution of the differences in multiplicities between
bins and to examine the scaling behavior of its moments. 

Another difficulty associated with the problem of linking multiplicity
fluctuations to the dynamics of phase transition is that the
particles are integrated over the entire duration when quarks turn
into hadrons, while the system undergoes an expansion.  In a
second-order PT the fluctuations in hadronization can be large
relative to the average, but that average is over a short period near
$T_c$ where hadronization is not robust.  When integrated over the
whole history of the nuclear collision process, such fluctuations may
well be averaged out, leaving no discernible effect at the detector,
which collects all the particles produced in an event.  This problem is
present even if there is no thermalization of the hadrons in the final
state, which we shall assume in order to focus our investigation here. 
Our present method is to apply the GL theory in increments of time
when hadrons are produced near $T_c$, and to integrate the
production process over the entire duration to yield the measurable
multiplicity fluctuations.  Our aim is to show that with appropriate
care in treating the moments certain scaling behavior persists and
characterizes the dynamics of phase transition.

Our result reveals a new scaling exponent $\gamma$, different
from the one, $\nu$, found in Refs. \cite{1,2}.  It is not a revision of
$\nu$ but a new exponent, since different quantities are investigated. 
Independent of the theoretical considerations underlying this work,
the proposed moments can be determined experimentally.  Nuclear
data should be analyzed in the way suggested, even if PT is not an
issue.  If the scaling behavior is found, but the scaling exponent does
not agree with the predicted 
$\gamma$, it would not only imply that there has not been a PT
of the GL type, but also present a numerical objective for a successful
hadronization model to attain for heavy-ion collisions.

\section{Multiplicity Difference Correlators}

To overcome the problem of high multiplicity per bin in heavy-ion
collisions we introduce the factorial moments that we shall refer to as
multiplicity difference correlators (MDC).  They are a form of hybrid
of factorial correlators \cite{9} and wavelets \cite{13}.  MDC are not as
elaborate as either one of the two separately, but are simpler
combinations of the two, possessing the virtues of both.

Let us start with a brief look at the application of wavelet analysis to
multiparticle physics \cite{13,14} and astrophysics \cite{15}.  Given
any event, the structure of a spatial pattern, exemplified here by a
one-dimensional function $g(x)$, can be analyzed by a multiresolution
decomposition, using the basis functions $\psi^H_{jk}(x)$ such that the
wavelet coefficients are
\begin{eqnarray}
W_{jk} = \int^L_0 dx \ \psi^H_{jk}(x)\ g(x) \quad ,
\label{2.1}
\end{eqnarray}
where $L$ is the domain of $g(x)$ being analyzed.  This is like a
Fourier transform, except that instead of the exponential factor
$\mbox{exp}(in \pi x/L)$, one uses the Haar wavelets
\begin{eqnarray}
 \psi^H_{jk}(x)\equiv \psi^H (2^j x - k) \quad ,
\label{2.2}
\end{eqnarray}
where
\begin{eqnarray}
 \psi^H(x)=\left\{\begin{array}{rl}
1, &0\leq x < 1/2\\
-1,&1/2 \leq x < 1\\
0, &\mbox{elsewhere} \quad .
\end{array}
\right.  
\label{2.3}
\end{eqnarray}
The basis function $\psi^H_{jk}(x)$ has a scale index $j$ and a
translation index $k$, which enable the wavelet analysis to identify
the location and scale of an arbitrary fluctuation of $g(x)$ in terms of
$W_{jk}$.

While (\ref{2.1}) and other transforms like it are very powerful and
have various virtues, such as their invertibility, they offer more than
what is needed at this point in our search for a measure of the
multiplicity fluctuations in heavy-ion collisions that can convey the
signature of phase transition.  A complete set of $W_{jk}$ records all
the information in the spatial pattern in an event, whereas we look
for an efficient way of capturing some general features averaged over
all events.

The Haar wavelets, however, contain some features that we regard as
important ingredients for an improvement of the usual factorial
moments and correlators \cite{9,10,17}.  For a fixed set of the indices
$j$ and $k$, we have
\begin{eqnarray}
\psi^H_{jk}(x)= \hspace{-.5cm}
&\mbox{}\ \ +1, \quad &\mbox{for}\, x\in x_+=\left\{x\,
|\, k/2^j\leq x < \left(k+{1\over 2}\right)/2^j\right\}\nonumber \\  
&\mbox{}-1, &\mbox{for}\, x\in x_-=\left\{x\,|\, \left(k + {1\over
2}\right)/2^j\leq x < (k+ 1)/2^j\right\}
\label{2.4}
\end{eqnarray}
so
\begin{eqnarray}
W_{jk} = n_+ - n_- \, , \hspace{2cm} n_{\pm} = \int_{x_{\pm}} dx \, g(x)
\quad.
\label{2.5}
\end{eqnarray}
It is the difference between the contributions in the neighboring bins,
$x_+$ and $x_-$.  In some wavelet analyses moments that involve sums
over $k$ are studied to reveal scaling behaviors in $j$
\cite{13}-\cite{15}.  That can be done for one event, sometimes referred
to as horizontal analysis.  For our purpose in what follows, we prefer to
emphasize first the vertical analysis; i.\ e., we average over all events for
fixed bins.  For multiparticle production $g(x)$ would be the particle
density and $n_{\pm}$ the multiplicities in the bins $x_{\pm}$.  Studying
the difference $n_+ - n_-$ of neighboring multiplicities is a way to
overcome the problem of high multiplicity per bin without abandoning
the focus on multiplicity fluctuations.

Once we consider multiplicity differences there is no reason why we
should restrict the two bins to only the neighboring ones.  Let $\Delta$
be the distance between two bins, each of size $\delta$, and let $n_1$
and $n_2$ be the multiplicities in those bins in a given event.  Define
$m$ to be the multiplicity difference, $m = n_1 - n_2$.  We shall be
interested in the distribution $Q_m (\Delta,\delta)$ after sampling
over many events at fixed $\Delta$ and $\delta$.  The factorial
moments we shall study are the MDC
\begin{eqnarray}
{\cal F}_q = {f_q  \over  f^q_1}, \quad f_q = \sum ^{\infty}_{m=q} m
(m-1) \cdots (m-q+1) Q_m \quad .
\label{2.6}
\end{eqnarray}
They are not the same as the Bia\l as-Peschanski correlators, which
are normalized products of factorial moments of multiplicities in two
bins, averaged over all events, \cite{9}
\begin{eqnarray}
F_{q_1q_2} = {\left<n_1\left(n_1-1\right) \cdots \left(n_1 - q_1
+1\right)n_2 \left(n_2 - 1\right)
\cdots \left(n_2 - q_2 + 1\right)\right>
\over  \left<n_1\right>^{q_1} \left<n_2\right>^{q_2}} \quad ,
\label{2.7}
\end{eqnarray}
but the two are similar to the extent that $n_1$ and $n_2$ are the
multiplicities in the two bins separated by $\Delta$.  $F_{q_1q_2}$
have been found to depend on $\Delta$, but not on $\delta$ \cite{10},
but ${\cal F}_q$ as defined in (\ref{2.6}) will depend on both.  ${\cal
F}_q$ differ from $F_q$ defined in (\ref{1}) in that they are the
moments of the multiplicity difference distribution (MDD)
$Q_m(\Delta,\delta)$, which is a generalization of the usual
multiplicity distribution in a way that incorporates the virtues of
both wavelets and correlators.

Before we enter into the theoretical description of the MDD $Q_m$ in
the following sections, we note that the experimental determination
of ${\cal F}_q$ for hadronic and nuclear collisions should be
performed independent of theory.  Their dependences on $\Delta$ and
$\delta$ will pose a challenge to any model of such collisions.  Our
concern in this paper is to make a theoretical prediction of what
should be observed when there is a quark-hadron PT.  But if there is
no PT, the properties of  ${\cal F}_q$ will remain as valuable
features of multiparticle production that a good model of soft
interaction must explain.

\section{Statistical and Dynamical Fluctuations}

Although our aim is to study the nature of the fluctuations due to
quark-hadron PT in heavy-ion collisions, we begin with a formulation
that is more generally valid for any hadronic or nuclear collisions. 
The multiplicities in the two bins discussed in the preceding section
fluctuate both statistically and dynamically.  Focusing on just the
statistical part first, and using, as usual, the Poisson distribution ${\cal
P}_{n_i}$ for it, we have the statistical MDD
\begin{eqnarray}
P_m \left(s_1, s_2\right) = \sum _{n_1,n_2}{\cal P}_{n_1}(s_1)
{\cal P}_{n_2}(s_2) \delta_{m-n_1 + n_2} \quad ,
\label{3.1}
\end{eqnarray}
where $s_1$ and $s_2$ are the average multiplicities in the two bins,
and 
\begin{eqnarray}
{\cal P}_{n_i}(s_i) = {1  \over  n_i!} s_i^{n_i} e^{- s_i} \quad .
\label{3.2}
\end{eqnarray}
Note that, unlike the usual multiplicity, $m$ can be both positive and
negative.  In fact, the sums in (\ref{3.1}) can be analytically
performed, yielding
\begin{eqnarray}
P_m \left(s_1, s_2\right) = \left(s_1/ s_2\right)^{m/2}I_m
\left(2\sqrt{s_1 s_2}\right) e^{-s_1- s_2}
 \quad ,
\label{3.3}
\end{eqnarray}
where $I_m$ is the modified Bessel function.

In the following we shall not assign any intrinsic properties to the
two bins and consider only the absolute difference between their
multiplicities, i.\ e.\,
\begin{eqnarray}
m = \left| n_1 - n_2\right| \quad .
\label{3.4}
\end{eqnarray} 
Since $I_m$ is symmetric under $m \leftrightarrow -m$, we have
from (\ref{3.3}) with $m \geq 0$
\begin{eqnarray}
P_m \left(s_1, s_2\right) = \cosh\left({m \over 2}\mbox{ln}{s_1 \over
s_2}\right)I_m
\left(2\sqrt{s_1 s_2}\right) e^{-s_1- s_2} \left(2 - \delta _{m0}\right)
 \quad .
\label{3.5}
\end{eqnarray}
There is a discontinuity at $m = 0$ because for all $m > 0$ there is a
reflection of $-m$ in (\ref{3.3}) to $+m$.  $P_m$ is properly
normalized to $\sum ^{\infty}_{m=0}P_m = 1$.

If $s_1$ and $s_2$ are large, then ${\cal P}_{n_i}(s_i)$ may be well
approximated by Gaussian distributions, and $P_m$ would also be
Gaussian with a width proportional to $\left(s_1s_2\right)^{1/4}$.  It
is because of this reduced width that we consider MDD:  as discussed
in Sec. 1, we need smaller values of $m$ to render lower-order
moments effective in measuring the fluctuations.  If $s_1$ and
$s_2$ are small, then $P_m$ becomes Poissonian also.  For that reason
we shall consider factorial moments of $P_m$, since the statistical
fluctuations can thereby be filtered out \cite{9}.  In both experimental
analysis and theoretical consideration the bin width $\delta$ is to be
varied so that $s_i$ will range over both large and small values;
hence, no approximation of $P_m$ will be made.

We now introduce the dynamical component of the fluctuations. 
Denoting it by $D \left(s_1, s_2, \Delta, \delta \right)$ we have for the
observable MDD
\begin{eqnarray}
Q_m(\Delta, \delta) = \int ds_1ds_2 P_m \left(s_1, s_2\right)  D
\left(s_1, s_2, \Delta, \delta \right) \quad .
\label{3.6}
\end{eqnarray}
In essence this is a double Poisson transform of the dynamical
$D$, which is a generalization of the formalism for photon
counting in quantum optics, later adapted by BP for particle
production \cite{9}.  It is this distribution
$Q_m$ that we have proposed to study by use of the MDC
${\cal F}_q$, defined in (\ref{2.6}).  Our aim is to examine the scaling
properties of ${\cal F}_q$ and extract universal features that are
characteristic of the dynamics of the problem.

There are many directions in which one can pursue from here.  For the
dynamical distribution $D$ one can consider the mathematical
$\alpha$ model \cite{18}, or the more physical models such as the
Fritiof model \cite{19}, VENUS \cite{20} and ECCO \cite{21}. 
Alternatively, one may want to emphasize the large-scale space-time
structure by considering the small $\Delta$ behavior in an
interferometry type of analysis \cite{22}.  For us in this paper we
want to consider the opposite, namely:  the large $\Delta$ behavior
where the global size of the particle emitting volume is unimportant. 
The usual short-range correlation in rapidity in low-$p_T$
multiparticle production \cite{10} would also be not important, if
$\Delta$ is sufficiently large, but not large enough to cause kinematical
constraint.  Our purpose is to go to a region where fluctuations due to
the dynamics of PT are the only ones that need to be taken into
account.

More specifically, we shall identify $D\left(s_1, s_2, \Delta, \delta
\right)$ with the Boltzmann factor, $\mbox{exp}(-F)$, in the
Ginzburg-Landau theory, in which $s_1$ and $s_2$ will be functions of
 the order parameter.  In fact, for two identical bins at large $\Delta$
apart in two regions of the expanding system that have similar spatial
and temporal properties, we may set $s_1 = s_2$, ignore $\Delta$ and
rewrite (\ref{3.6}) as
\begin{eqnarray}
Q_m (\delta) = \int ds \, P_m (s) D (s, \delta) \quad.
\label{3.7}
\end{eqnarray}
There remains a complication arising from the property that PT occurs
over an extended period of time and that the detected hadrons are
the result of an integration over that period.  That is the subject we
turn to in the next section.

\section{Hadronization in Ginzburg-Landau Theory}

The conventional view of the physical system in a heavy-ion collision
at very high energy is that a cylinder of locally thermalized partons
expands as a fluid, mainly in the longitudinal direction, but also in the
radial direction at a slower rate.  If the colliding nuclei are massive
enough, and the incident energy high enough, the temperature in the
interior of the cylinder may be higher than the critical temperature
$T_c$ for quark-hadron PT, which we shall assume to be second
order.  Due to the transverse expansion the temperature $T$
decreases with increasing radius, at least initially and for the most
part of the lifetime of the system.  Thus hadronization takes place
mainly on the surface of the cylinder where $T
\approx T_c$.  Being a second-order PT there is no mixed phase
where quarks and hadrons coexist.  We assume that there is no
thermalized hadronic phase surrounding the partonic cylinder, so the
hadrons formed on the surface move in free flow to the detector. 
With these simplifying assumptions, which are not unrealistic at
extremely high collision energies, we can then focus on the issue of
relating the hadronization process on the surface to the hadron
multiplicity collected by the detector.

Consider first just one bin, which occupies $\delta \eta\,  \delta
\varphi$ in pseudorapidity $\eta$ and azimuthal angle $\varphi$.  We
define $\delta^2$ to be its area.  Such an area selected by an analyst of
the data corresponds to a similar area on the surface of the cylinder.  Let
the hadronization time be $t_h$, which is of order 1 fm/c: it is the
average time for the formation of one hadron on the surface.  During that
time we use the GL description of PT to specify the probability
$D(s,\delta)$ that $s$ hadrons are created in the area
$\delta^2$.  The GL free energy, being time independent, does not track
the time evolution of the system.  It is also not equipped to describe the
spatial variations on the surface, nor need it be.  As a mean field theory,
it is concerned with the probability for PT near $T_c$ as a function of the
order parameter $\phi$.  Let the two-dimensional coordinates on the
surface be labeled by $z$, then the GL free energy is \cite{23}
\begin{eqnarray}
F[\phi] = \int_{\delta ^2}dz \left[a \left|\phi(z)\right|^2 +
b\left|\phi(z)\right|^4  + c\left|\partial\phi/\partial z \right|^2 \right]
\quad ,
\label{4.1}
\end{eqnarray}
where $a$, $b$, and $c$ are GL parameters that depend on $T$.  For
hadronization within the bin of size $\delta^2$ it is only necessary to
integrate $z$ over that area.  For $T \stackrel{<}{\sim} T_c$ the GL
theory requires that $a < 0$, and $b > 0$.  We have found in
\cite{1,3} that for small bins the third term in (\ref{4.1}) does not
have any significant effect on the multiplicity fluctuations, so we shall
set $c = 0$, as it has been done in all previous work
\cite{1}-\cite{8}.  Furthermore, we make the approximation that
$\phi(z)$ is constant in $\delta^2$ so that
\begin{eqnarray}
F[\phi] = \delta ^2 \left(a \left|\phi\right|^2 +
b\left|\phi\right|^4   \right)
\quad ,
\label{4.2}
\end{eqnarray}
Assuming that this is valid for any local area on the cylindrical
surface, the same free energy is to be used for both bins of $P_m$ in
(\ref{3.6}), resulting in the same average multiplicity $s_1$ and
$s_2$.  We now must specify the relationship between $s=s_1=s_2$
and $\left|\phi\right|^2$.  

As discussed extensively in \cite{2}, the square of the order parameter
$\left|\phi\right|^2$ is the hadron density $\rho$, which, in the
absence of fluctuations around the minimum of (\ref{4.2}) at
$\left|\phi_0\right|^2$, is zero for $T > T_c$ but positive for $T < T_c$. 
Thus the average multiplicity in a bin is
\begin{eqnarray}
\bar{n}_0 = \delta ^2 \left|\phi\right|^2 \quad .
\label{4.4}
\end{eqnarray}
It should be borne in mind, however, that this is the average
multiplicity during the hadronization time $t_h$, when the system is
momentarily regarded as static, and the GL consideration is applied to
describe the formation of hadrons from quarks at $T
\stackrel{<}{\sim} T_c$.  It is not the average multiplicity in $\delta
^2$ registered by the detector, since the experimental measurement
integrates the hadronization process over the entire lifetime ${\cal T}$
of the whole parton system, during which partons are continuously
converted to hadrons.  Thus the measured average multiplicity in
$\delta^2$ is 
\begin{eqnarray}
s = \delta ^2 \int^{\cal T}_0 dt\left|\phi(t)\right|^2 \quad ,
\label{4.5}
\end{eqnarray}
where $\phi(t)$ is formally the time-dependent order parameter,
which has to be varied in a functional-integral description of
$Q_m(\delta)$.  Indeed, $s$ in (\ref{4.5}) is the average multiplicity of
$P_m(s)$ in (\ref{3.7}), but $\bar{n}_0$ in (\ref{4.4}) is the relevant
average multiplicity for the GL description during $t_h$.  Thus to
adapt the formalism of the previous sections to the PT process in
heavy-ion collisions, it is necessary to modify (\ref{3.7}) into a
functional integral
\begin{eqnarray}
Q_m (\delta, \tau) = Z^{-1}\int {\cal D}\phi \ P_m \left(\delta^2 \tau
\left|\phi \right|^2\right)  \mbox{exp}\left[-\delta^2 \left( a\left|\phi
\right|^2 + b\left|\phi \right|^4\right)
\right]
\quad ,
\label{4.6}
\end{eqnarray}
where ${\cal D}\phi = \pi d\left|\phi\right|^2$, $Z = \int {\cal D}\phi\,
\mbox{exp}\, [-F(\phi)]$ and the integral in (\ref{4.5}) has been
discretized into $\tau = {\cal T}/t_h$ segments.  In each segment
$\phi$ is spatially and temporally constant in $\delta^2$ and is itself
integrated over the whole complex plane.  For collisions of large nuclei
$\tau$ may be a large number.  Herein lies the crux of the problem:  PT
at $T \stackrel{<}{\sim} T_c$ gives sparingly few hadrons within
$\delta ^2$ in any time interval around $t_h$, but the detected number
in a bin at the end of the collision process is roughly $\tau$ times as
many.  Our aim to find some universal feature of the problem that is
essentially independent of $a$, $b$, $\delta$ and $\tau$.

\section{Scaling Behavior}

We now proceed to determine the multiplicity difference correlator MDC,
which in the present case of large $\Delta$ has no dependence on
$\Delta$ and is just the normalized factorial moments ${\cal F}_q$ of the
MDD
$Q_m$ defined in (\ref{2.6}).  As we have mentioned in general terms
earlier, an important reason for studying $Q_m$ instead of the usual
multiplicity distribution $P_n$ in a single bin is the largeness of
$\delta^2 \tau
\left|\phi \right|^2$ in (\ref{4.6}), when $\tau$ is large.  In that case the
bin multiplicity $n$ is high, so factorial moments of low orders are
ineffective in extracting genuine correlations.  Multiplicity difference
$m$ is on the average proportional to $\sqrt{\tau}$, thus enabling ${\cal
F}_q$ to be more effective at low $q$.

Setting $s_1 = s_2 = s$ in (\ref{3.5}), with 
\begin{eqnarray}
s = \delta ^2 \tau \left|\phi \right|^2 \quad ,
\label{5.1}
\end{eqnarray}
we have the distribution $P_m$ in (\ref{4.6})
\begin{eqnarray}
P_m(s) = \left(2 - \delta_{m0}\right) I_m (2s) e ^{-2s} \quad ,
\label{5.2}
\end{eqnarray}
where $m \geq 0$.  Let us simplify (\ref{4.6}) by using the variable
$u^2 = \delta^2 \, b \left|\phi \right|^4$, getting 
\begin{eqnarray}
Q_m (\tau, x) = Z^{-1} \int ^\infty_0 du \ P_m  (\tau x u) \, e ^{xu - u^2}
\quad ,
\label{5.3}
\end{eqnarray}
where
\begin{eqnarray}
Z =  \int ^\infty_0 du \  e ^{xu - u^2} \quad, \qquad x = \left|a
\right|\delta /\sqrt{b} \quad .
\label{5.4}
\end{eqnarray}
For notational economy $\tau$ has been redefined as 
\begin{eqnarray}
\tau = {{\cal T} \over t_h \left| a\right|}
\label{5.5}
\end{eqnarray}
It is therefore the $\tau$ defined earlier in units of $\left| a\right|$.  It
makes physical sense because $\left| a\right|$ is a measure of how
readily hadronization can take place in the GL theory.  The minimum of
$F [\phi]$ in (\ref{4.2}) is at $\left| \phi_0\right|^2 = \left| a\right|/2b$,
for $a < 0$, so there would be virtually no hadron condensates apart from
fluctuations, if $\left| a\right| \rightarrow 0$, resulting in $\tau
\rightarrow \infty$.  Thus the expanding parton system must drive the
surface temperature to below $T_c$, making $a$ sufficiently negative
and $\left| \phi_0\right|^2$ large enough to produce hadrons at a rate
just such as to carry away the necessary energy to maintain the
hydrodynamical flow with $T < T_c$ at the surface.  The beauty of the GL
approach is that all the complications of the hydrodynamics of the
problem are hidden in a few parameters, which would be in the final
answer [like $\tau$ and $x$ in (\ref{5.3})] unless we can find
observables that are insensitive to them.

To find such observables we now calculate the normalized factorial
moments ${\cal F}_q$ of $Q_m$ using (\ref{2.6}).  We first fix $\tau$ and
examine ${\cal F}_q$ as a function of $x$, which is proportional to
$\delta$.  In Fig. 1 we show how $\mbox{ln} {\cal F}_q$ depends on
$\log_2 x$ for $\tau = 10$ and $2 \leq q \leq 6$.  No simple behavior can
be ascribed to the rising and falling of the curves.  It should be
noticed that the increase of ${\cal F}_q$ with increasing $\delta$ is
opposite to the usual behavior of the single-bin factorial moments $F_q$
\cite{1,3,7}, which increase with decreasing $\delta$.  Thus ${\cal F}_q$
do not have the intermittency behavior of BP \cite{9}.

The similarity of the dependences of ${\cal F}_q$ on $x$ for the various
$q$ values shown in Fig. 1 suggest that we should examine, as in \cite{1},
the dependences of ${\cal F}_q$ on ${\cal F}_2$ for $3 \leq q \leq 6$. 
That is shown in Fig. 2, where the dots are the values of ${\cal F}_q$ for
$\log _2 x = i/2$ with $i$ being integers in the range $-12 \leq i \leq
4$.  The straightlines are fits of the linear portions.  Clearly, ${\cal F}_q$
exhibit the power-law behavior
\begin{eqnarray}
{\cal F}_q \propto {\cal F}_2^{\beta _q} \quad ,
\label{5.6}
\end{eqnarray} 
which we shall call $F$-scaling.  It is a scaling behavior that is
independent of $\delta$ and $b$.  The dependence on $a$ is so far
unknown, since the calculation is done at a fixed $\tau$.

From the slopes of the straight lines in Fig. 2 we show in Fig. 3 the
dependence of $\beta_q$ on $q$.  It can be well fitted by the formula
\begin{eqnarray}
\beta _q = (q - 1)^{\gamma} \quad ,
\label{5.7}
\end{eqnarray}
where $\gamma = 1.099$.  We use the symbol $\gamma$ here for the
scaling  exponent, instead of $\nu$, which we have used previously for a
similarly defined quantity [as in (\ref{5.7})] for $F_q$.  For comparison,
we recall that \cite{1}
\begin{eqnarray}
\nu = 1.304 \quad .
\label{5.8}
\end{eqnarray}
Thus the scaling exponent $\gamma$ for MDC is significantly lower. 
Lower value of that exponent means larger fluctuations. 

So far the result is for $\tau = 10$ only.  To see the dependences on
$\tau$, we have repeated the calculation for a range of $\tau$ values. 
Scaling behavior as in (\ref{5.6}) has been found in each case, and
(\ref{5.7}) is also well satisfied.  Fig. 4 shows the dependence of
$\gamma$ on $\tau$.  Evidently, it is nearly constant for $3 < \tau < 30$
with the value
\begin{eqnarray}
\gamma = 1.09 \pm 0.02 \quad .
\label{5.9}
\end{eqnarray}
Thus, the result of this study is embodied in just one number,
$\gamma$.  It is independent of $a$, $b$, $\delta$, and $\tau$, provided
that $a$ is negative to allow hadronization.

The universality of $\gamma$ is remarkable and should be checked
experimentally.  If a signature of quark-hadron PT depends on the
details of the heavy-ion collisions, such as nuclear sizes, collision energy,
transverse energy, etc., even after they have passed the thresholds for
the creation of quark-gluon plasma, such a signature is likely to be
sensitive to the theoretical model used.  Here $\gamma$ is independent
of such details; it depends only on the validity of the GL description of PT
for the present problem.  The MDD $Q_m$ can readily be measured
experimentally, and the moments  ${\cal F}_q$ directly determined as
functions of bin size.  If the scaling behavior (\ref{5.6}), supplemented by
(\ref{5.7}), is satisfied with $\gamma \simeq 1.1$, then we may interpret
the system as having undergone a second-order PT describable by the
GL theory.

In the absence of PT we may regard the two bins separated by a large
$\Delta$ to be totally uncorrelated.  In that case $Q_m (\delta)$ can be
identified with $P_m[s(\delta)]$, where, if we vary $s$ in the range $1 <
s < 100$, the corresponding ${\cal F}_q$ without PT can be calculated
directly.  Again, $F$-scaling is found, satisfying (\ref{5.6}) and
(\ref{5.7}), but this time with 
\begin{eqnarray}
\gamma = 1.33 \pm 0.02 \qquad \qquad \mbox{(no PT)}\quad .
\label{5.910}
\end{eqnarray}
This value is sufficiently separated from that of (\ref{5.9}) derived
for PT so that phenomenological distinquishability between the two
cases should be quite feasible.

\section{Conclusion}

In this work we have solved a number of problems that have obstructed
the study of multiplicity fluctuations in heavy-ion collisions as a means
of finding signatures of quark-hadron phase transition.  One problem is
the large multiplicities even in small bins, for which the usual factorial
moments $F_q$ fail to reveal distinctive features for $3 \leq q
\leq 6$, since events with large fluctuations are submerged by generic
events.  That is mainly an experimental problem where the analysis of
the data cannot be pushed to the regions $\left<n\right>_{\delta} < q$. 
Another problem of more theoretical nature is that the application of the
Ginzburg-Landau theory of PT needs special tailoring for a system whose
lifetime is finite, but long compared to the transition time for individual
hadrons, and whose observables are integrated over that time.  We have
overcome both of these problems by showing the effectiveness of
studying the fluctuations of multiplicity differences.

We have started with wavelet analysis and found that it generates more
information than can easily be filtered to yield a succinct signature of
PT.  However, we extracted a simple feature of the wavelets and
considered the MDD, $Q_m$, involving two bins separated by a distance
$\Delta$.  Although $\Delta$ can be any value, we have considered only
large $\Delta$ in order to apply the simplest description of PT by the GL
theory.  The time integration problem of the detectable multiplicities is
handled at the expense of an extra parameter $\tau$, which cannot be
specified without a hydrodynamical model lying outside the scope of this
treatment.  The goal has then been to find a measure of PT that is
independent of the unknown parameters in the problem.

That goals was achieved by the discovery that the factorial moments
${\cal F}_q$ of $Q_m$ satisfy a scaling behavior that is characterized by
a number $\gamma \simeq 1.1$.  It is independent of the details of the
dynamics, except that a PT occurs in a way describable by the the GL
theory.  Thus $\gamma$ is a universal constant for the problem.  We
call $\gamma$ a scaling exponent, but it has absolutely no connection
with the critical exponents of the conventional critical phenomena. 
There is no need to know the temperature, which is not measurable, or
the critical $T_c$.  The moments ${\cal F}_q$ can directly be determined
from the data, and can therefore be checked for $F$-scaling,
independent of any theoretical input.  Whatever experimental value
the exponent $\gamma$ turns out to be would be of great interest, since
a value different from 1.1 would demand an explanation.  If it
approaches 1.1 as energy or nuclear size is increased, it would suggest
an approach to the condition required for the formation of quark-gluon
plasma. 

Experimentally, it should be easy to vary $\Delta$ and check not only
how
${\cal F}_q$ themselves depend on $\Delta$, but also whether and how
the scaling behavior is affected.  The $\Delta$ dependence has not been
investigated here.  Theoretically, there are many other challenges that
also await undertaking.  A verification of $\gamma = 1.1$ will
undoubtedly stimulate an upsurge of interest in the study of multiplicity
fluctuations in particle and nuclear collisions.
\begin{center}
\subsubsection*{Acknowledgment}
\end{center}

Discussion and communication with Z. Cao, L.Z. Fang and P. Lipa on
wavelets during the early phase of this work have been beneficial.  This
work was supported in part by U.S. Department of Energy under Grant No.
DE-FG03-96ER40972.

\vspace*{2.5cm}
\begin{center}
\section*{Figure Captions}
\end{center}
\begin{description}

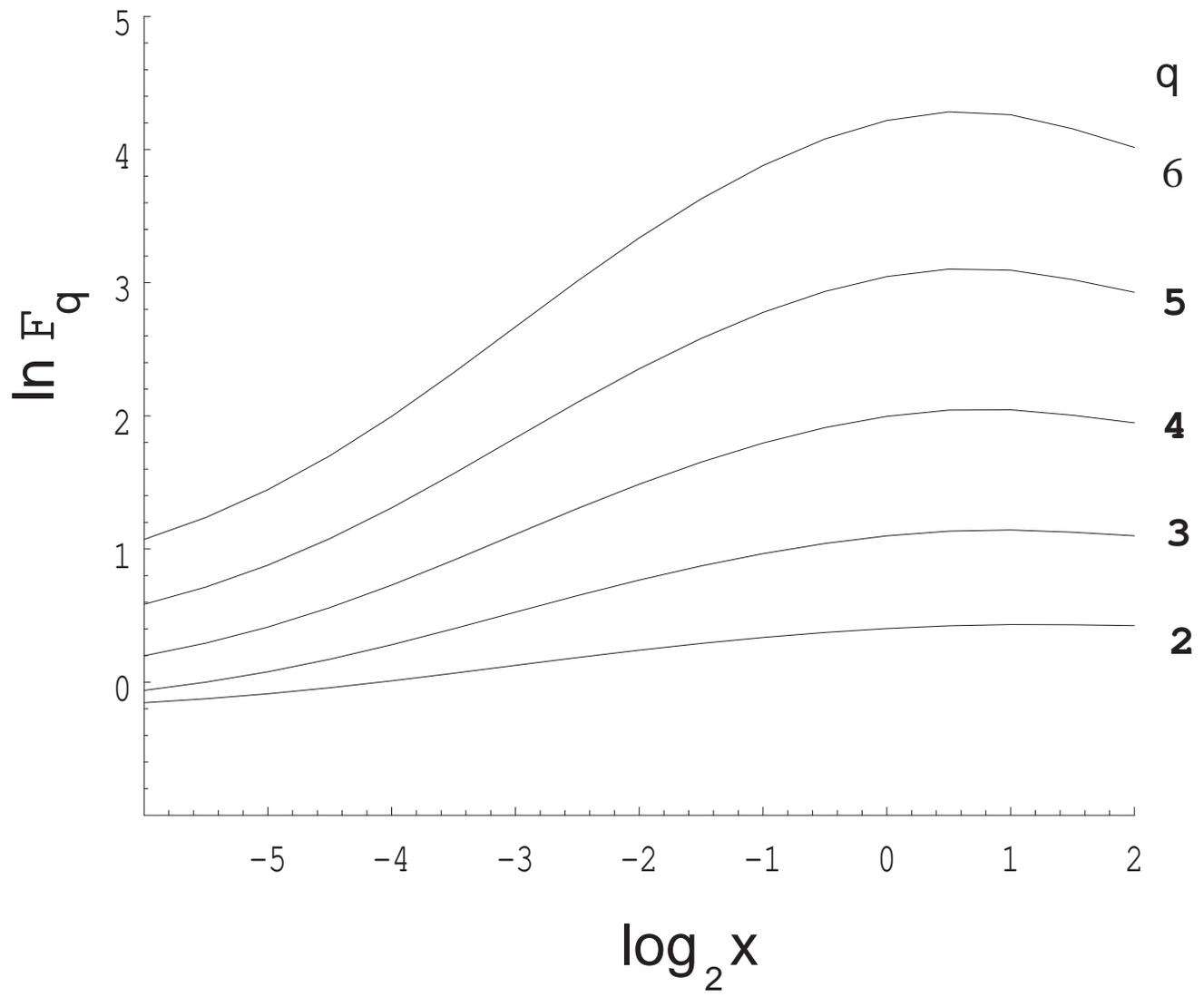
\item[Fig.\ 1]\quad Factorial moments ${\cal F}_q$ versus $x$ in log-log
plot for various values of $q$ at $\tau=10$.

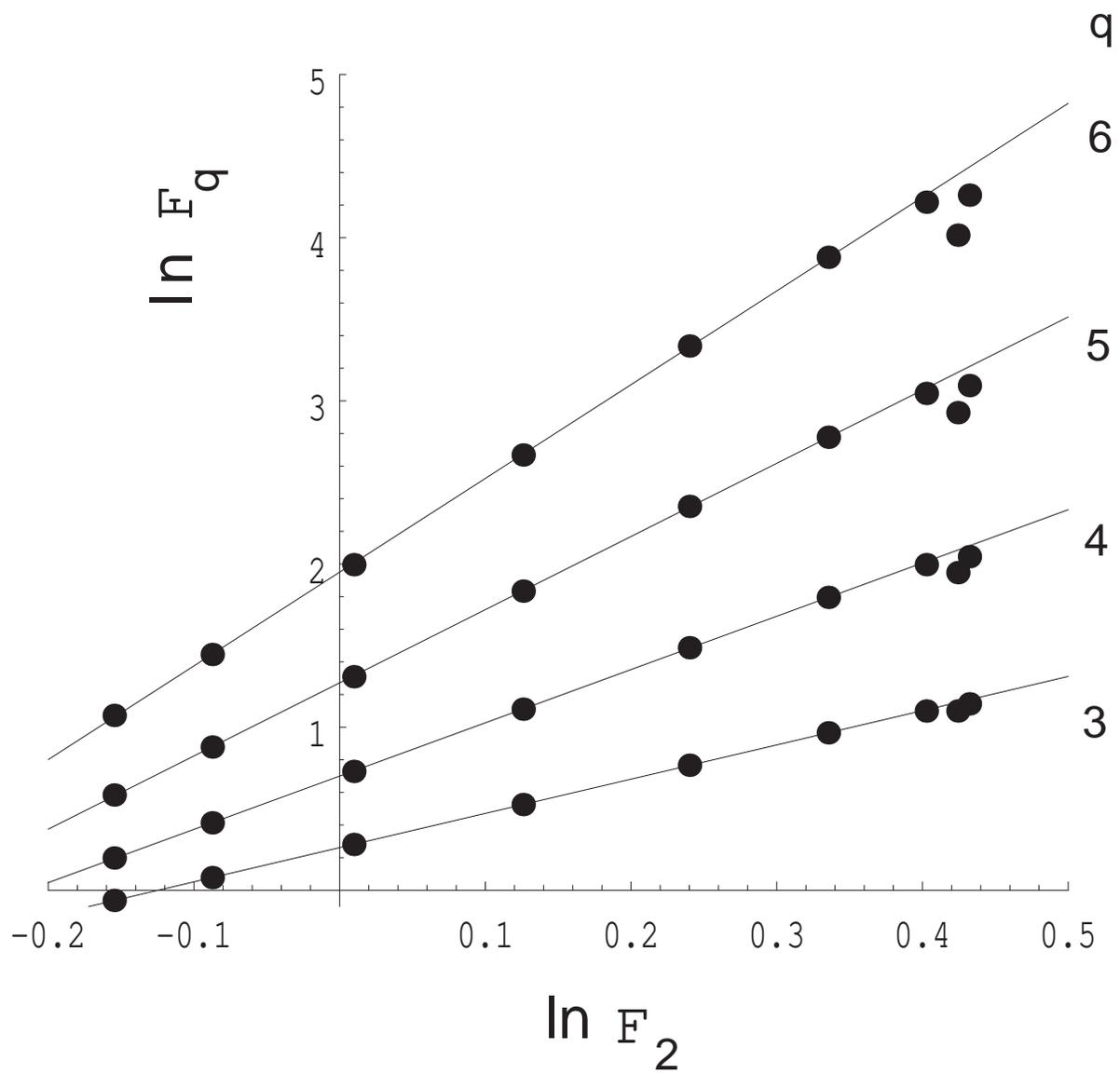
\item[Fig.\ 2]\quad Factorial moments ${\cal F}_q$ versus ${\cal F}_2$ in
log-log plot for various values of $q$ at $\tau=10$. Dots are calculated
values at $x=2^{i/2}$ for $-12\leq i\leq 4$. Straight lines are fits of the
linear portions of the dots.

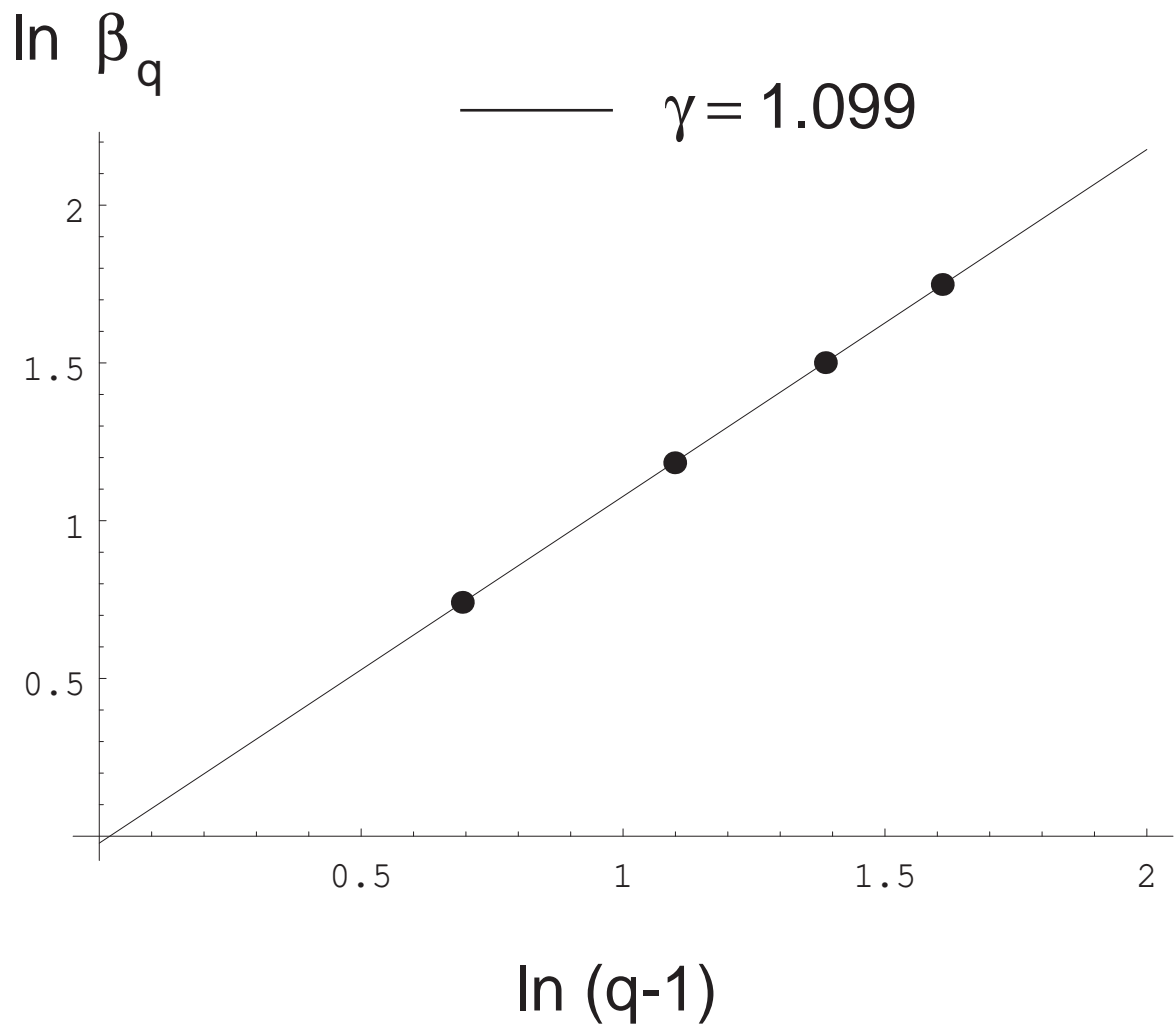
\item[Fig.\ 3]\quad Dots are the slopes of the straight lines in Fig.\ 2,
plotted against $q-1$ in log-log plot. The straight line is the best fit of
the dots, whose slope is $\gamma$, defined in Eq.\ (28).

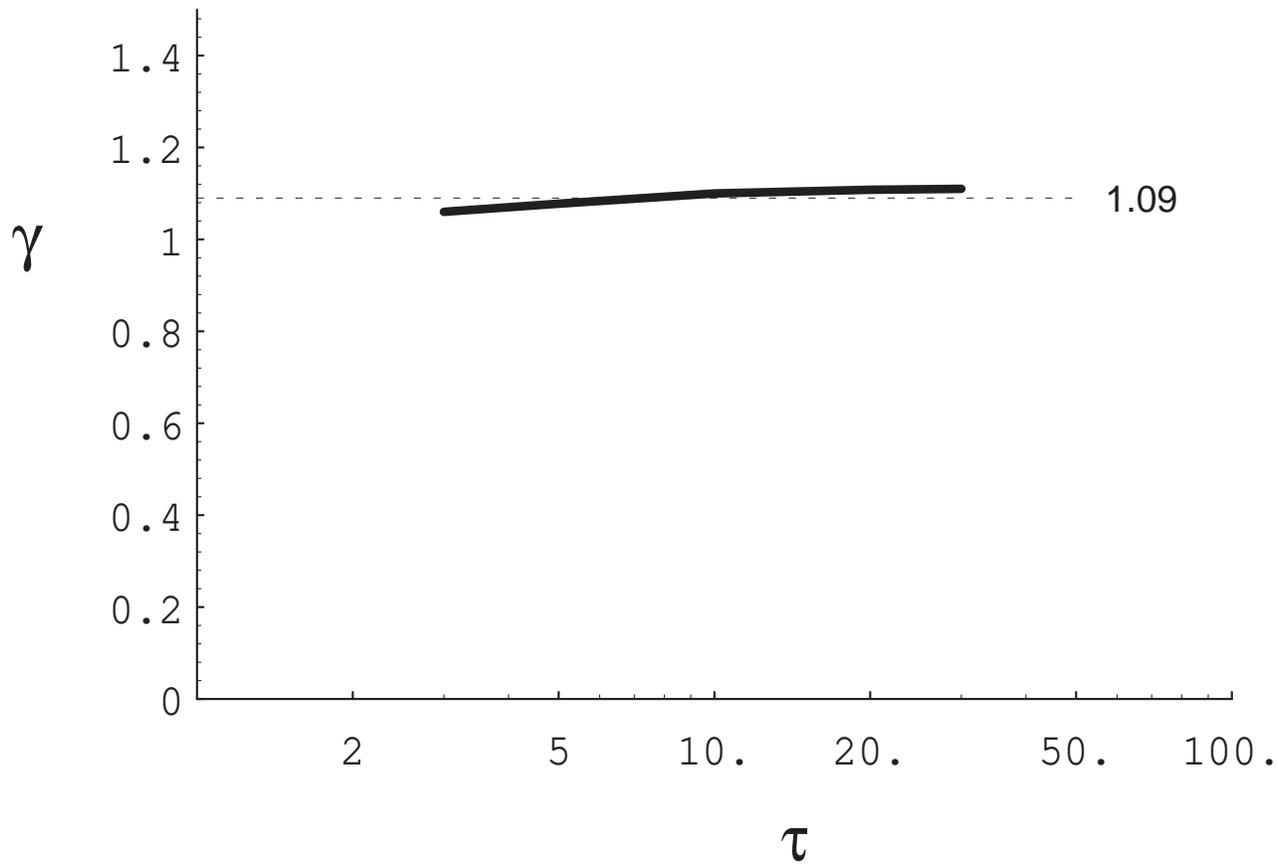
\item[Fig.\ 4]\quad Solid line shows the dependence of $\gamma$ on
$\tau$; dotted line is at $\gamma=1.09$ to guide the eye.

\end{description}

\newpage
\begin{figure}
\centerline{\epsfbox{fig1m.epsf}}
\end{figure}
\newpage
\begin{figure}
\centerline{\epsfbox{fig2m.epsf}}
\end{figure}
\newpage
\begin{figure}
\centerline{\epsfbox{fig3m.epsf}}
\end{figure}
\newpage
\begin{figure}
\centerline{\epsfbox{fig4m.epsf}}
\end{figure}

\end{document}